\documentclass[a4paper,preprint]{aastex631}
\usepackage{tabularx, booktabs}
\usepackage{graphicx}
\usepackage{float}
\usepackage{subfigure}
\usepackage{rotating} 
\usepackage{threeparttable}
\usepackage{array}
\usepackage{longtable}
\usepackage{hyperref}
\usepackage{lineno}

\newcommand{\Ha}{H$\alpha$}

\newcommand{\OII}{[{O}~{\scriptsize {II}}]}

\newcommand{\SII}{[{S}~{\scriptsize {II}}]}

\newcommand{\hi}{{H}{\sc{i}}}

\begin{document}

\shorttitle{Impact of gas accretion on \hi\ distribution}
\shortauthors{Zhang et al.}

\title{The impact of external gas accretion on the distribution of \hi\ gas in galaxies}

\author[0009-0005-7716-3144]{Qianhan Zhang}
\author[0009-0005-9342-9125]{Min Bao}\thanks{Qianhan Zhang and Min Bao share first authorship.}
\author[0000-0003-3226-031X]{Yanmei Chen}\thanks{E-mail: \url{chenym@nju.edu.cn}.}
\affiliation{School of Astronomy and Space Science, Nanjing University, Nanjing 210023, China}
\affiliation{Key Laboratory of Modern Astronomy and Astrophysics, Nanjing University, Ministry of Education, Nanjing 210023, China}
\author[0000-0002-9066-370X]{Niankun Yu}
\affiliation{Max Planck Institute for Radio Astronomy, Auf dem H$\ddot{u}$gel\ 69, 53121\ Bonn, Germany}
\affiliation{National Astronomical Observatories, Chinese Academy of Sciences, Beijing, 100101, China} 
\affiliation{Key Laboratory of Radio Astronomy and Technology, Chinese Academy of Sciences, Beijing, 100101, China}

\author[0000-0002-8614-6275]{Yong Shi}
\affiliation{School of Astronomy and Space Science, Nanjing University, Nanjing 210023, China}
\affiliation{Key Laboratory of Modern Astronomy and Astrophysics, Nanjing University, Ministry of Education, Nanjing 210023, China}
\author[0000-0001-6947-5846]{Luis C. Ho}
\affil{Kavli Institute for Astronomy and Astrophysics, Peking University, Beijing 100871, China}
\affil{Department of Astronomy, School of Physics, Peking University, Beijing 100871, China}

\begin{abstract}
Using the data from Mapping Nearby Galaxies at Apache Point Observatory (MaNGA) and {\hi}-MaNGA surveys, we build a sample of 37 gas-star misaligned galaxies with robust {\hi} detections, which are believed to have undergone external gas accretion processes. Both star-forming (SF) and quiescent (QS) misaligned galaxies exhibit narrower {\hi} line widths compared to their gas-star aligned controls. The {\hi} profiles of SF misaligned galaxies tend to be single-peaked, displaying a slightly higher fraction of single-peaked shape compared to their aligned controls. The  QS misaligned galaxies exhibit prominently single-peaked {\hi} profiles, while their aligned controls show distinct double-horned profiles. The shape of {\hi} profiles is expected to change with the {\hi} surface density radial gradients through external gas accretion --- the interaction between the accreted gas and the pre-existing gas leads to the re-distribution of angular momentum and induces gas inflow. It suggests that the progenitors of SF misaligned galaxies are central {\hi}-enriched, in this case, the shape of {\hi} profiles is insensitive to the further increase of central {\hi} surface density. The progenitors of QS misaligned galaxies are central {\hi}-deficient, hence the transition from central {\hi}-deficient to {\hi}-enriched surface density leads to significantly more single-peaked {\hi} profiles.

\end{abstract}


\section{Introduction} \label{sec:intro}

According to conservation law of angular momentum (AM), stars ought to share the same direction of AM as gas if the evolution of a galaxy is dominated by internal processes. However, several decades ago, \citet{Galletta1987} found the AM direction of the ionized gas and stellar disks being misaligned with each other in a SB0 galaxy NGC~4564. With the developments of long slit and integral field spectroscopic observations, galaxies with kinematically misaligned gas and stellar components (misaligned galaxies for short) were found to be common in the quiescent (QS) sequence with old stellar populations, in which the misaligned fraction stands 20\%--50\% \citep{Sarzi2006,Davis2011,Barrera-Ballesteros2015}. Meanwhile, this fraction turned out to be much lower (2\%--5\%) in the star-forming (SF) main sequence \citep{Chen2016,Jin2016,Bryant2019}.

The gas-star misaligned phenomenon is believed to originate from external processes, including gas accretion and/or galaxy mergers \citep{Corsini2014}. \cite{L'Huillier2012} used a multi-zoom simulation based on TreeSPH code to analyze the mass assembly of 530 galaxies, revealing that smooth accretion dominates the assembly (77\% of total mass), while mergers contribute obviously less (23\%). \cite{Lagos2015} identified a similar trend in their simulations of QS misaligned galaxies: mergers alone produced misalignment in only 2\% QS galaxies, while adding gas accretion processes increased this fraction to 46\% --- in good agreement with observational results. Furthermore, using the Mapping Nearby Galaxies at Apache Point Observatory (MaNGA) observations, \cite{Li2021} uncovered that the fraction of misaligned galaxies with merger remnant features exhibits insignificant increase compared to their control sample, illustrating that mergers are not the primary processes for the misaligned phenomenon. These results of simulations and observations support a consistent scenario that the formation of misaligned galaxies is dominated by external gas accretion.

Based on the MaNGA observations, \cite{Bao2025} found that the gas spins in misaligned galaxies tend to be perpendicular to the large-scale filaments, demonstrating that the misaligned gas is accreted from the large-scale structure. However, the impact of accretion on the gas content in misaligned galaxies is still not completely understood. On the one hand, gas accretion leads to an immediate increase in the gas content inside these galaxies. The collision between the pre-existing and accreted misaligned gas can redistribute angular momentum, and trigger gas inflow. On the other hand, the inflowing gas provides material for star formation and black hole activity in the galactic center \citep{Lu2021}. The feedback of these processes drives gas outflow, which can eventually reduces the gas content inside misaligned galaxies \citep{Duckworth2020a, Duckworth2020b}. Through regulating the rates of gas inhalation and consumption, external gas accretion plays a critical role in the subsequent evolution of misaligned galaxies.

Previous studies found that the impact of gas accretion on the evolution of misaligned galaxies can be different in SF sequence and QS sequence. Based on a sample of SF misaligned galaxies selected from the early date release of MaNGA survey, both \cite{Chen2016} and \cite{Jin2016} found that the stellar populations in the central region of SF misaligned galaxies are younger than that in their outskirts, suggesting that the interaction between external accreted and pre-existing gas re-distributes the AM of gas component, leading to the gas inflow which triggers star formation in the central region. Different from SF misaligned galaxies, the stellar populations of QS misaligned galaxies show similar negative gradients as that of aligned controls, suggesting external gas accretion does not trigger the formation of new stars \citep{Jin2016,Xu2022}. Furthermore, \cite{Zhou2022} studied the global properties of the misaligned galaxies, finding larger asymmetry in ionized gas velocity fields in the misaligned galaxies than their aligned controls in both SF sequence and QS sequence. They also analyzed the \hi\ detection rate and molecular gas mass fraction, finding that they are significantly lower in misaligned galaxies than their control samples in both sequences.

Cold gas, the fuel of star formation in galaxies, plays a critical role in galaxy evolution \citep{Kennicutt2012}. 21~cm neutral hydrogen (\hi) gas is one of the main components of cold gas, and it has been proven to be a sensitive indicator of gas interactions due to its extended spatial distribution \citep{Kornreich2000, Reichard2008, Wang2024}. The extended \hi\ disk can be easily distorted by external perturbation, such as tidal interaction, ram pressure stripping and gas accretion. \cite{Yun1994} discovered the interaction debris on M81 group through the spatially resolved \hi\ observations, which is absent in their optical image. Although the single-dish observations are much cheaper than the spatially resolved ones, it still can provide plentiful information on the \hi\ gas. For instance, the integrated flux of \hi\ spectrum is a good indicator of \hi\ gas mass, the line width and shape of \hi\ spectrum can give clues on the spatial and velocity distribution of \hi\ gas.

In this study, we build a sample of 37 misaligned galaxies with robust \hi\ observations. By comparing their \hi\ profiles with gas-star aligned control galaxies (aligned controls for short), we explore the impact of external gas accretion on the distribution of \hi\ gas in galaxies. In Section \ref{sec:Data analyses}, we introduce the data analysis and sample selection. We stack the \hi\ spectra for misaligned galaxies and aligned controls, respectively, and make a comparison between them in Section \ref{sec:spectral stacking}. In Section \ref{sec:spectral measurements}, we quantify the difference in \hi\ profiles between misaligned galaxies and aligned controls using the curve-of-growth (CoG) method. Based on these results, we discuss the impact of gas accretion in Section \ref{sec:impact}. Section \ref{sec:conclusion} is the conclusion.

\section{Data and Sample Selection} \label{sec:Data analyses}

\subsection{MaNGA survey} \label{subsec:MaNGA}

MaNGA is one of the three core programs of the fourth generation of Sloan Digital Sky Survey (SDSS-IV; \citealt{Bundy2015}). It used the 2.5-m Sloan Foundation Telescope at Apache Point Observatory \citep{Gunn2006} to observe 10,010 nearby galaxies \citep{Abdurro'uf2022}. These galaxies cover a flat distribution of stellar mass in $\log(M_\ast / M_\odot) \sim [9, 11]$, and redshift interval of $z \sim [0.01, 0.15]$ \citep{Blanton2017}. Two dual-channel BOSS spectrographs provide simultaneous wavelength coverage of [3600, 10,300]~$\rm \AA$, which enable the measurements of all prominent emission lines from \OII $\lambda\lambda$3726,3729 to \SII $\lambda\lambda$6717,6731. The wavelength calibration accuracy is $\sim 5~\rm km\ s^{-1}$. The spectral resolution is $R \sim$ 2000.

The data products in the optical band are obtained from the final data release of MaNGA survey. The global properties of MaNGA galaxies in terms of stellar mass, redshift and photometric axial ratio in r-band ($q=b/a$) are obtained from the NASA-Sloan Atlas catalog (NSA\footnote{http://nsatlas.org}, \citealt{Blanton2011}). The global stellar mass is estimated from spectral energy distribution (SED) fitting by the \texttt{kcorrect} package \citep{Blanton&Roweis2007} with BC03 single stellar population model \citep{Bruzual&Charlot2003} and \citet{Chabrier2003} initial mass function.
The axial ratio is converted into inclination angle following:
\begin{equation}
\sin\ i=\sqrt{\frac{1-q^2}{1-q_0^2}}, \label{con:inclination}
\end{equation}
where the intrinsic axial ratio ($q_0$) is set to be 0.2. Furthermore, the MaNGA Data Reduction Pipeline (DRP; \citealt{Law2016}) offers sky-subtracted and flux-calibrated 3D spectra. We stack the spectra with signal-to-noise (S/N) per pixel larger than 2, and measure the global 4000~\AA~break (D$_n4000$) from the stacked spectrum of each galaxy. The MaNGA Data Analysis Pipeline (DAP; \citealt{Westfall2019}) offers measurements of spatially resolved properties, such as stellar and gas kinematics.

\subsection{\hi-MaNGA survey} \label{subsec:HI-MaNGA}

\hi-MaNGA\footnote{https://greenbankobservatory.org/science/gbt-surveys/hi-manga} \citep{Masters2019,Stark2021} is a follow-up survey of \hi\ observations for the MaNGA galaxies with $z \leq 0.05$. The data products of \hi\ spectra are obtained from the third data release of the \hi-MaNGA value-added catalog, including the integrated \hi\ flux ($F_{\rm HI}$), flux-weighted central velocities ($V_{\rm C}$) and $W_{\rm 50}$ defined as the line width at 50\% of the peak flux in the case of a single-peaked spectrum (or the maximum width at 50\% of the peak flux on either side of a double-horned spectrum). In this catalog, 1147 galaxies are detected by Arecibo telescope \citep{Haynes2018} with spectral resolution of $\rm \sim 5.5~ km~s^{-1}$, and 2030 galaxies are detected by Green Bank Telescope (GBT) with spectral resolution of $\rm \sim 5~km~s^{-1}$.

Given the large beam size of Arecibo telescope ($\sim 3.5^{\prime}$) and GBT ($\sim 9^{\prime}$), the \hi\ observations of MaNGA galaxies could be contaminated by companions. \cite{Witherspoon2024} found that some active galactic nuclei (AGN) in low-mass galaxies with \hi\ detection in the \hi-MaNGA catalog are actually \hi-poor, but contaminated by \hi-rich companions. \cite{Stark2021} marked the potential contaminated sources located within 1.5 times the half-power beamwidth of the primary target and at a similar redshift. Additionally, they estimated the probability that more than 20\% flux in this beam is contributed by companions ($P_{R > 0.2}$), where $R$ corresponds to the fraction of contribution to the total measured flux from the companions. \cite{Stark2021} suggested $P_{R > 0.2} < 0.1$ as a criteria for selecting galaxies whose contamination of \hi\ emission can be neglected.


\subsection{Sample selection} \label{subsec:sample selection}

\begin{figure}[h]
\centering
\includegraphics[width=\textwidth]{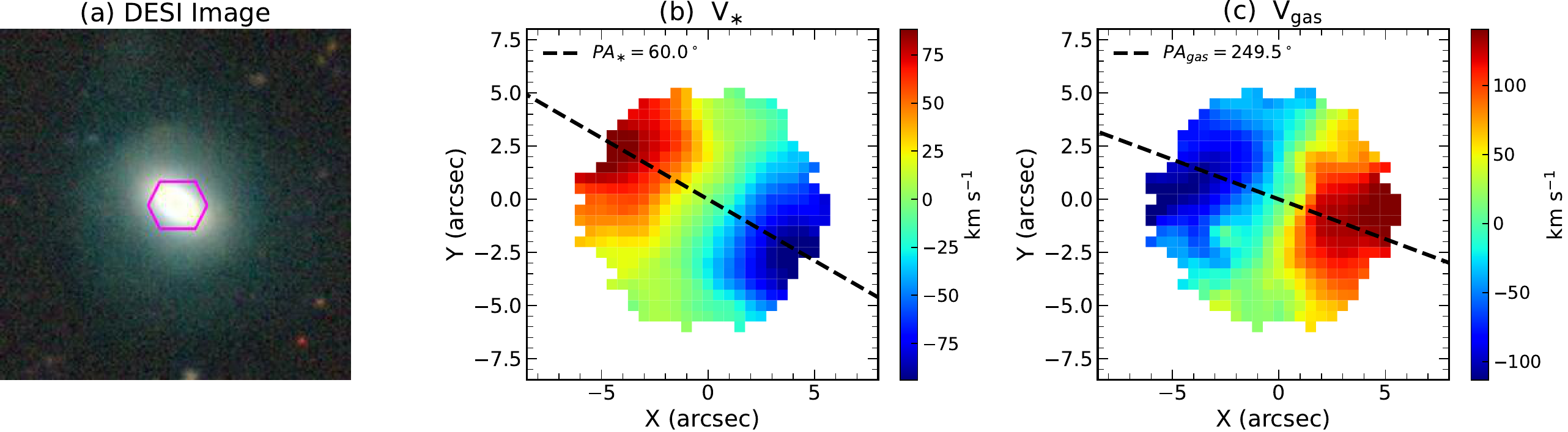}
\caption{An example of misaligned galaxies MaNGA 1-418253. (a) The DESI $g, r, z$ image. The purple hexagon shows the MaNGA bundle. (b) Stellar velocity field. The black dashed line shows the kinematic position angle the stellar disk. (c) Gas velocity field. The black dashed line shows the kinematic position angle of the gas disk. In panels (b) \& (c), the blue and red colors represent velocities approaching and receding from the galactic center. }
\label{f:case}
\end{figure}

Following the method of \cite{Zhou2022}, we build a sample of 496 misaligned galaxies from the final data release of MaNGA survey. Firstly, we collect 7459 emission-line galaxies with the signal-to-noise ratios (S/N) of \Ha\ emission lines larger than 3 for at least 10\% spaxels within $1.5\ R_e$. Secondly, we fit the kinematic position angle for gas ($\rm PA_{gas}$) and stellar ($\rm PA_\ast$) components using Python-based code \texttt{PAFIT} \citep{Krajnovic2006}. The position angle is defined as the counterclockwise angle between the north and the major axis of the velocity field on the receding side. The misaligned candidates are selected as $\Delta \rm PA \equiv \left| {\rm PA}_{gas}-{\rm PA}_\ast \right| \geq 30^\circ$, and $\rm PA_{error} \leq 20^\circ$ is required for robust position angle measurements, where $\rm PA_{error}$ is the 1$\sigma$ error of PA. Finally, we visually inspect these candidates to remove ongoing mergers and galaxy pairs located within the same MaNGA bundle. Figure~\ref{f:case} displays an example of misaligned galaxies. Figure~\ref{f:case}(a) shows the DESI image of $g, r, z$ bands with the MaNGA bundle shown as a purple hexagon. Figures~\ref{f:case}(b) and~\ref{f:case}(c) display the stellar and gas velocity fields, respectively. The black dashed line marks the PA for the corresponding component, with PA values labeled in the top-left corner of each panel. It is clear that this is a counter-rotator with $\Delta \rm PA \sim 180 ^\circ$ between stellar and gas components, an extreme subclass in gas-star misalignment.

We cross-match the misaligned sample with the \hi-MaNGA value-added catalog \citep{Stark2021}, obtaining \hi\ spectra for 102 misaligned galaxies. The S/N of each \hi\ spectrum can be defined as 
\begin{equation}
 S/N = \frac{F_{HI}}{\sigma\delta V\sqrt{N_{chan}}},
\label{con:S/N}
\end{equation}
where $\sigma$ is the root-mean-square (rms) of a spectrum, which represents the noise level (in unit of Jy), $\delta V$ is the spectral resolution in unit of $\rm km~s^{-1}$, and $N_{chan}$ is the channel number of spectrum. By analyzing 10,000 mocked \hi\ spectra with varying S/N, \cite{Yu2022} demonstrated that the statistic analysis of spectra with $\rm S/N>5$ is reliable. Therefore, we exclude 7 misaligned galaxies with \hi\ spectral $\rm S/N \leq 5$. Moreover, contamination from companions can lead to inaccurate measurements of \hi\ profiles. We further exclude 58 misaligned galaxies whose \hi\ emission is contaminated by companions as: (1) $\rm W_{50} > 500~km~s^{-1}$ or (2) $P_{R\ > 0.2} > 0.1$. Finally, a sample of 37 misaligned galaxies with robust \hi\ observations is selected for the following studies. We list MaNGA-ID as well as some basic parameters of the 37 misaligned galaxies in the Table~\ref{t:misaligned properties}.

\begin{figure}[h]
\centering
\includegraphics[width=0.9\textwidth]{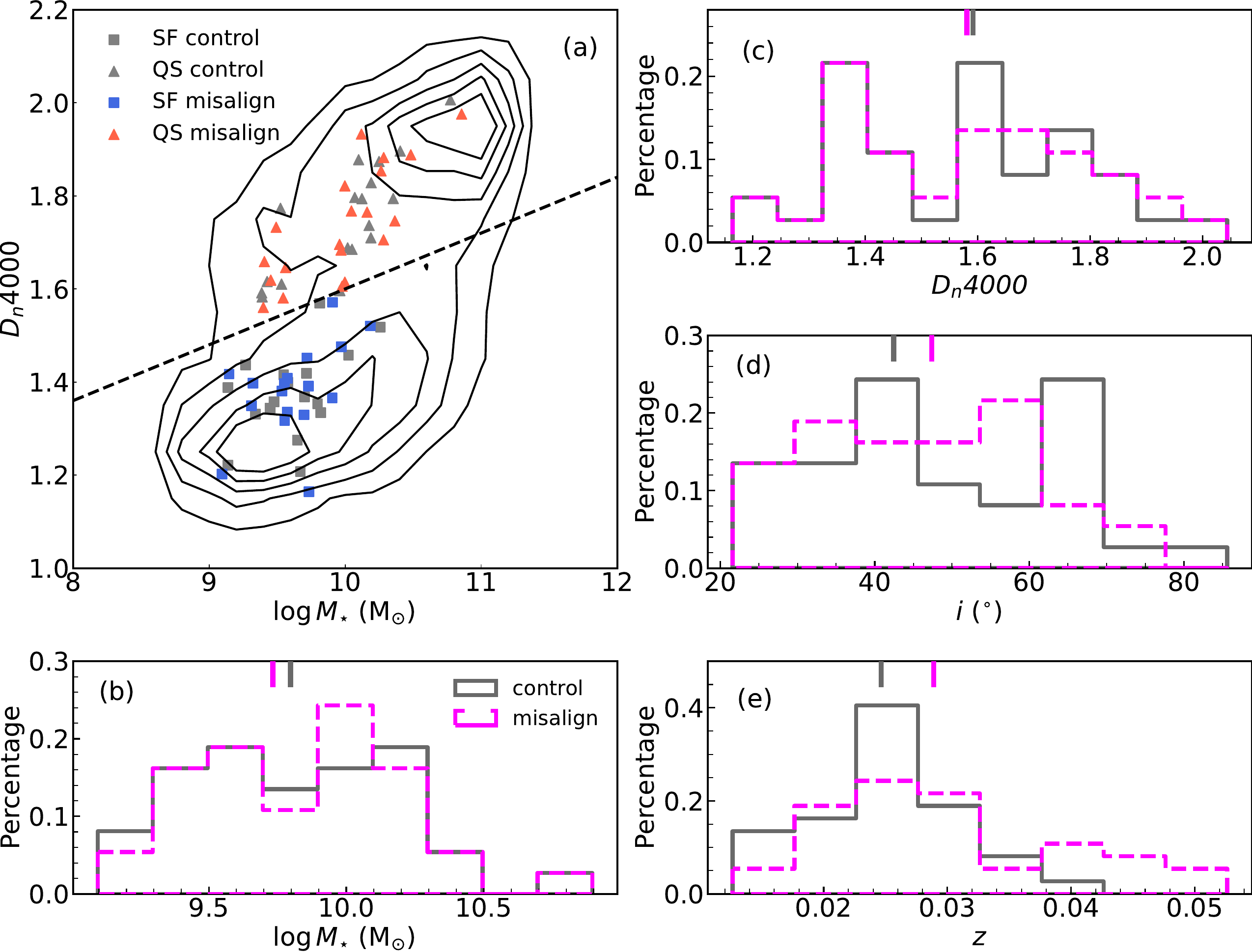}
\caption{The physical properties of misaligned galaxies and aligned controls. (a) The galaxy distribution in a two-dimensional plane of $M_\ast$ and D$_n4000$. The black contour shows the distribution of SDSS galaxies. The black dashed line separates galaxies into SF sequence and QS sequence. The blue and gray squares represent SF misaligned galaxies and aligned controls. The red and gray triangles represent QS misaligned galaxies and aligned controls. (b) Stellar mass distribution. (c) D$_n4000$ distribution. (d) Inclination angle distribution. (e) Redshift distribution. In panels (b), (c), (d) and (e), the pink and gray histograms represent the distributions of misaligned galaxies and aligned controls, respectively. The pink and gray bars on the top show the corresponding median values.}
\label{f:sample}
\end{figure}

Figure~\ref{f:sample}(a) displays the galaxy distribution in a two-dimensional plane of global $M_\ast$ and D$_n4000$. D$_n4000$ is strongly correlated with the light-weighted stellar population age, and sensitive to star formation in $\sim$Gyr timescale. The black contour shows the distribution of 10,010 galaxies from SDSS DR17, which presents two density peaks. The peak at the bottom-left with lower $M_\ast$ and D$_n4000$ corresponds to the SF sequence, while the peak at the top-right with higher $M_\ast$ and D$_n4000$ corresponds to the QS sequence. To compare the impact of gas accretion on SF and QS galaxies, we classify the misaligned galaxies into SF sequence and QS sequence with a cutoff showed as black dashed line. The blue squares and red triangles show the distributions of 17 SF and 20 QS misaligned galaxies, respectively. We apply $M_\ast$-D$_n4000$ rather than $M_\ast$-SFR relation to separate SF sequence and QS sequence, because: (1) different from SFR, D$_n4000$ can be measured consistently for SF and QS galaxies; (2) the interaction between pre-existing and accreted external gas can lead to higher SFR in SF misaligned galaxies \citep{Chen2016,Jin2016,Xu2022}, while it has a smaller impact on D$_n4000$ which indicates star formation in $\sim$Gyr timescale.

\begin{table}[thp]
\caption{Misaligned galaxies with robust \hi\ measurements}
\footnotesize
\setlength{\tabcolsep}{16pt} 
\resizebox{\textwidth}{!}{
\begin{threeparttable} 
\begin{tabularx}{\textwidth}{ccccccccccc}
\tableline
MaNGA-ID & Class & $z$ & {$\rm \log(M_\ast/M_\odot)$} & {D$_n4000$} & $i$ & {$\rm W_{50}$} & S/N \\
 & & & & &(deg) & (km/s) & \\
\text{(1)} & \text{(2)} & \text{(3)} &  \text{(4)} & \text{(5)} & \text{(6)} & \text{(7)} & \text{(8)}\\
\tableline
1-137890 & SF & 0.027 & 9.58 & 1.34 & 58.51 & 174.55 & 18.50 \\
1-138140 & SF & 0.047 & 10.18 & 1.52 & 30.29 & 138.48 & 7.77 \\
1-199775 & SF & 0.030 & 9.91 & 1.37 & 41.11 & 114.55 & 7.58 \\
1-547295 & SF & 0.041 & 9.53 & 1.38 & 45.59 & 105.59 & 11.41 \\
1-38887 & SF & 0.024 & 9.31 & 1.35 & 61.65 & 165.08 & 9.84 \\
1-207984 & SF & 0.038 & 9.55 & 1.40 & 37.89 & 187.68 & 16.25 \\
1-246517 & SF & 0.019 & 9.15 & 1.42 & 55.01 & 147.06 & 11.68 \\
1-94690 & SF & 0.031 & 9.91 & 1.57 & 60.66 & 292.68 & 10.10 \\
1-272473 & SF & 0.044 & 9.97 & 1.48 & 66.56 & NaN & 10.39 \\
1-51810 & SF & 0.021 & 9.72 & 1.45 & 22.63 & 142.08 & 42.89 \\
1-153038 & SF & 0.018 & 9.10 & 1.20 & 32.89 & 70.96 & 10.69 \\
1-109275 & SF & 0.044 & 9.73 & 1.39 & 49.78 & 218.69 & 12.94 \\
1-153901 & SF & 0.040 & 9.70 & 1.33 & 48.04 & 124.59 & 8.68 \\
1-24055 & SF & 0.030 & 9.32 & 1.40 & 55.64 & 78.66 & 9.81 \\
1-118363 & SF & 0.026 & 9.57 & 1.41 & 70.95 & NaN & 13.53 \\
1-587628 & SF & 0.019 & 9.73 & 1.16 & 28.11 & NaN & 18.47 \\
1-386932 & SF & 0.026 & 9.56 & 1.32 & 51.00 & NaN & 5.88 \\
1-339061 & QS & 0.020 & 10.00 & 1.61 & 53.37 & 336.72 & 25.11 \\
1-338746 & QS & 0.032 & 10.16 & 1.77 & 26.78 & 153.24 & 7.40 \\
1-235530 & QS & 0.027 & 10.00 & 1.82 & 26.96 & 239.09 & 15.64 \\
1-93378 & QS & 0.049 & 10.26 & 1.85 & 36.00 & 230.07 & 9.63 \\
1-94228 & QS & 0.049 & 10.28 & 1.88 & 30.80 & 36.58 & 5.14 \\
1-561039 & QS & 0.026 & 9.97 & 1.68 & 71.04 & 195.44 & 26.36 \\
1-178824 & QS & 0.013 & 9.45 & 1.62 & 55.27 & 45.62 & 9.24 \\
1-180080 & QS & 0.030 & 10.28 & 1.71 & 31.78 & 496.26 & 18.31 \\
1-38543 & QS & 0.023 & 10.05 & 1.77 & 58.19 & 313.24 & 6.50 \\
1-71956 & QS & 0.040 & 9.54 & 1.58 & 32.81 & 82.24 & 9.86 \\
1-456984 & QS & 0.018 & 10.86 & 1.98 & 21.89 & NaN & 8.36 \\
1-174947 & QS & 0.033 & 10.37 & 1.75 & 60.20 & 400.39 & 17.77 \\
1-188177 & QS & 0.027 & 9.98 & 1.61 & 43.73 & 164.21 & 22.45 \\
1-246484 & QS & 0.030 & 9.41 & 1.66 & 54.02 & 158.37 & 9.82 \\
1-462966 & QS & 0.032 & 10.48 & 1.89 & 47.52 & 253.48 & 14.23 \\
1-153127 & QS & 0.018 & 9.49 & 1.73 & 37.68 & 298.95 & 55.02 \\
1-189376 & QS & 0.019 & 9.56 & 1.65 & 33.11 & 105.78 & 6.33 \\
1-319646 & QS & 0.029 & 9.96 & 1.70 & 64.66 & 122.42 & 17.80 \\
1-152010 & QS & 0.028 & 9.40 & 1.56 & 47.37 & 149.40 & 7.00 \\
1-198125 & QS & 0.036 & 10.12 & 1.93 & 43.38 & 378.41 & 9.81 \\
\tableline
\end{tabularx}
\begin{tablenotes}   
\footnotesize             
\item(1) MaNGA-ID; 
(2) The classification of the galaxy using stellar mass and $D_n4000$, including star-forming (SF) and quiescent (QS) sequence;
(3) Redshift;
(4) Stellar mass;
(5) Global 4000\AA\ break;
(6) The inclination angle of the galaxy calculated through Equation~\ref{con:inclination};
(7) Line width at 50\% of the peak flux in the case of a single-peaked spectrum (or the maximum width at 50\% of the peak flux on either side of a double-horned spectrum);
(8) The signal-to-noise ratio of the \hi\ spectrum.
\end{tablenotes} 
\end{threeparttable}}
\label{t:misaligned properties}
\end{table}

\subsection{Control sample} \label{subsec:control}

In order to understand the impact of external gas accretion on the distribution of \hi\ gas, we build an aligned control sample with $\Delta \rm PA < 30^{\circ}$, \hi\ spectral $\rm S/N > 5$, $W_{50} < 500~\rm km~s^{-1}$ and the contamination from companions can be neglected ($P_{R > 0.2} < 0.1$). For each misaligned galaxy, one aligned control galaxy is closely matched in three dimensional parameter space of stellar mass ($|\Delta\log {M_\ast}| < 0.1$), 4000~\AA~break ($|\Delta$D$_n4000| < 0.05$) and inclination angle ($|\Delta i| < 10 ^\circ$).
The motivation for choosing these parameters is: (1) stellar mass is the most fundamental property of a galaxy, and it is tightly related to various other physical properties; (2) the similar global D$_n4000$ ensures that misaligned galaxies and aligned controls have similar stellar populations with comparable light-weighted ages averaged over the past few Gyrs; and (3) the inclination of a galaxy can affect the observed \hi\ profile \citep{El-Badry2018}.

The gray squares in Figure~\ref{f:sample}(a) show the distribution of 17 SF aligned controls, and gray triangles show the distribution of 20 QS aligned controls. Figures~\ref{f:sample}(b),~\ref{f:sample}(c) and~\ref{f:sample}(d) display the distributions of $\log M_\ast$, D$_n4000$, and $i$ for misaligned galaxies (pink histogram) and their aligned controls (gray histogram), respectively. The pink and gray bars mark the median values for corresponding distributions. Figure~\ref{f:sample}(e) displays the distribution of $z$, color-coded in the same way as other panels. The difference in median redshifts between misaligned galaxies and aligned controls is less than 0.005.

\section{Individual \& stacking \hi\ profiles} \label{sec:spectral stacking}

\begin{figure}[h]
\centering
\includegraphics[width=\textwidth]{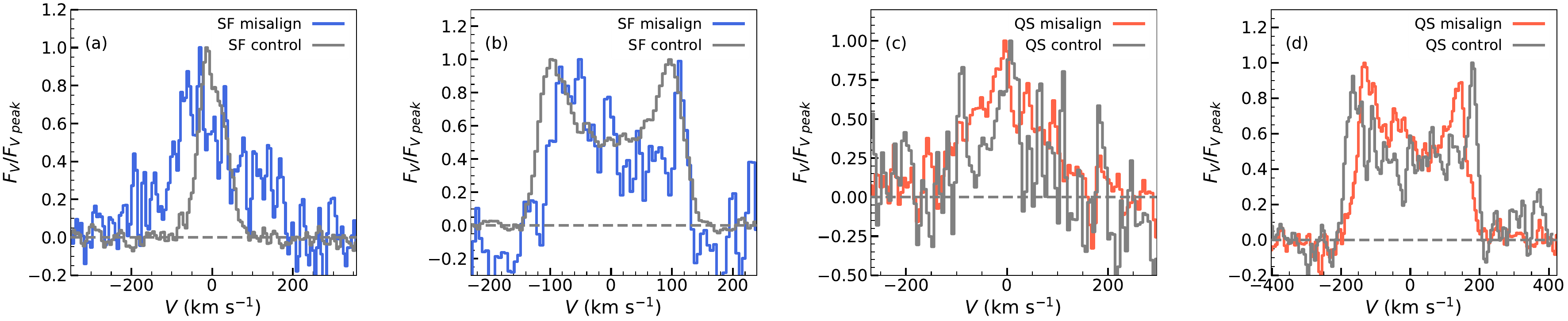}
\caption{Examples of \hi\ profiles. (a) Star-forming (SF) misaligned galaxy and aligned control with single-peaked \hi\ profiles; (b) SF misaligned galaxy and aligned control with double-horned \hi\ profiles; (c) Quiescent (QS) misaligned galaxy and aligned control with single-peaked \hi\ profiles; (d) QS misaligned galaxy and aligned control with double-horned \hi\ profiles. Blue (red) lines represent the SF (QS) misaligned galaxies and gray lines represent the aligned controls. Each profile is normalized by its peak flux.}
\label{f:examples}
\end{figure}

To obtain an intuitive sense of the \hi\ profiles in misaligned galaxies and their aligned controls, we first check the \hi\ profile of each individual galaxy. Figure~\ref{f:examples} show typical examples of \hi\ profile, for each sub-sample (SF misalign vs. control, QS misalign vs. control), one single-peaked \hi\ profile and one double-horned profile are provided. Each profile is normalized by its peak flux. Panels (a) and (b) show SF misaligned galaxies (blue) and their controls (gray). 11 out of 17 SF misaligned galaxies ($\sim65\%$) exhibit single-peaked profiles, while the single-peaked fraction in SF control is $\sim47\%$. Panels (c) and (d) show QS misaligned galaxies (red) and their controls (gray). 13 out of 20 QS misaligned galaxies (65\%) show single-peaked profile, while the fraction in QS control is only 25\%, which is 1.6 times lower than that of the QS misaligned galaxies.

\begin{figure}[h]
\centering
\includegraphics[width=0.9\textwidth]{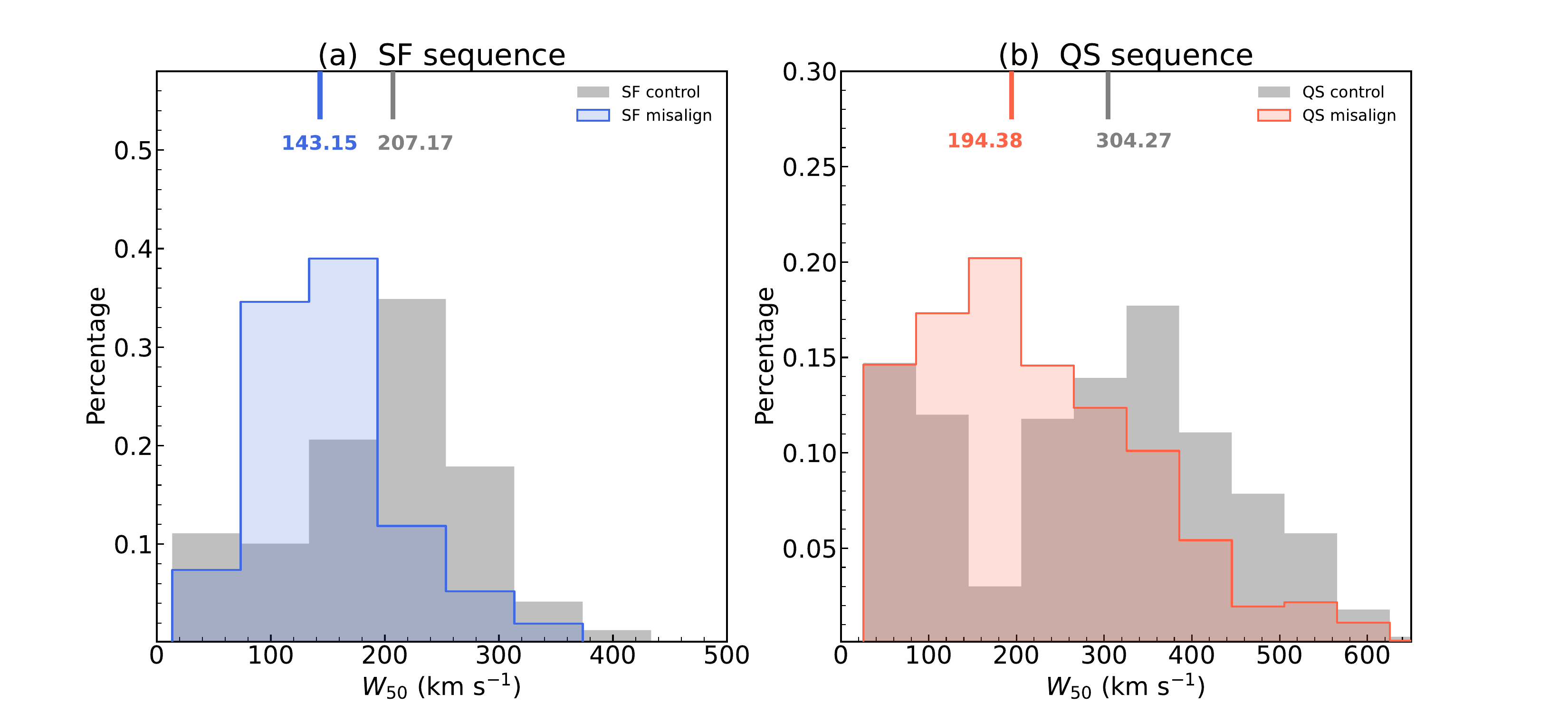}
\caption{The distributions of line widths ($\rm W_{50}$) for misaligned galaxies and aligned controls. (a) The blue and gray histograms show the line widths distribution for SF misaligned galaxies and their aligned controls, respectively. The blue and gray bars on the top mark the corresponding median values. (b) The red and gray histograms show the line widths distribution for QS misaligned galaxies and their aligned controls, respectively. The red and gray bars on the top mark the corresponding median values.}
\label{f:linewidth distribution}
\end{figure}

We further compared the \hi\ linewidth ($\rm W_{50}$) of these galaxies. It should be noted that, 32 galaxies have reliable $\rm W_{50}$ measurements from \hi-MaNGA catalog, while the others lack $\rm W_{50}$ measurement due to the non-smooth \hi\ profile. Figure~\ref{f:linewidth distribution} displays the $\rm W_{50}$ distributions for these 32 misaligned galaxies and their aligned controls. Panel (a) shows SF misaligned galaxies (blue) and their controls (gray). It is significant that the misaligned galaxies exhibit narrower $\rm W_{50}$ compared with their aligned controls. Panel (b) shows QS misaligned galaxies (red) and their controls (gray) with the difference of $\rm W_{50}$ in QS sequences being more obvious than that in SF sequences.


In addition, to provide a clear, visual illustration of the characteristic differences in \hi\ profiles between the misaligned galaxies and their controls, we compare the stacked \hi\ spectra between them. The process of \hi\ spectral stacking includes:

\noindent(1) subtracting the systemic velocity $V_{\rm C}$ to shift each baseline-subtracted \hi\ spectrum to the rest frame;

\noindent(2) extracting each \hi\ spectrum in a velocity range of $[-500, 500]~\rm km~s^{-1}$, and normalizing it by its peak flux;

\noindent(3) adjusting the channel widths to $\rm 5~km~s^{-1}$ through linear interpolation;

\noindent(4) ensuring that the integrated flux in $0<V<500~\rm km~s^{-1}$ is larger than that in $-500<V<0~\rm km~s^{-1}$, otherwise, mirroring the spectrum;

\noindent(5) stacking the \hi\ spectra following a weighting function:
\begin{equation}
 S_{stack}=\frac{\sum_{i=0}^{N}{S_iw_i}}{\sum_{i=0}^{N}w_i} \label{con:stacking}
\end{equation}
where $S_i$ is \hi\ spectrum of a galaxy after the preceding steps, $w_i=1/{rms_i}^2$ is the corresponding weighting factor, where $rms$ is the root mean square noise of the spectra \citep{Silvia2012}.

\begin{figure}[h]
\centering
\includegraphics[width=0.8\textwidth]{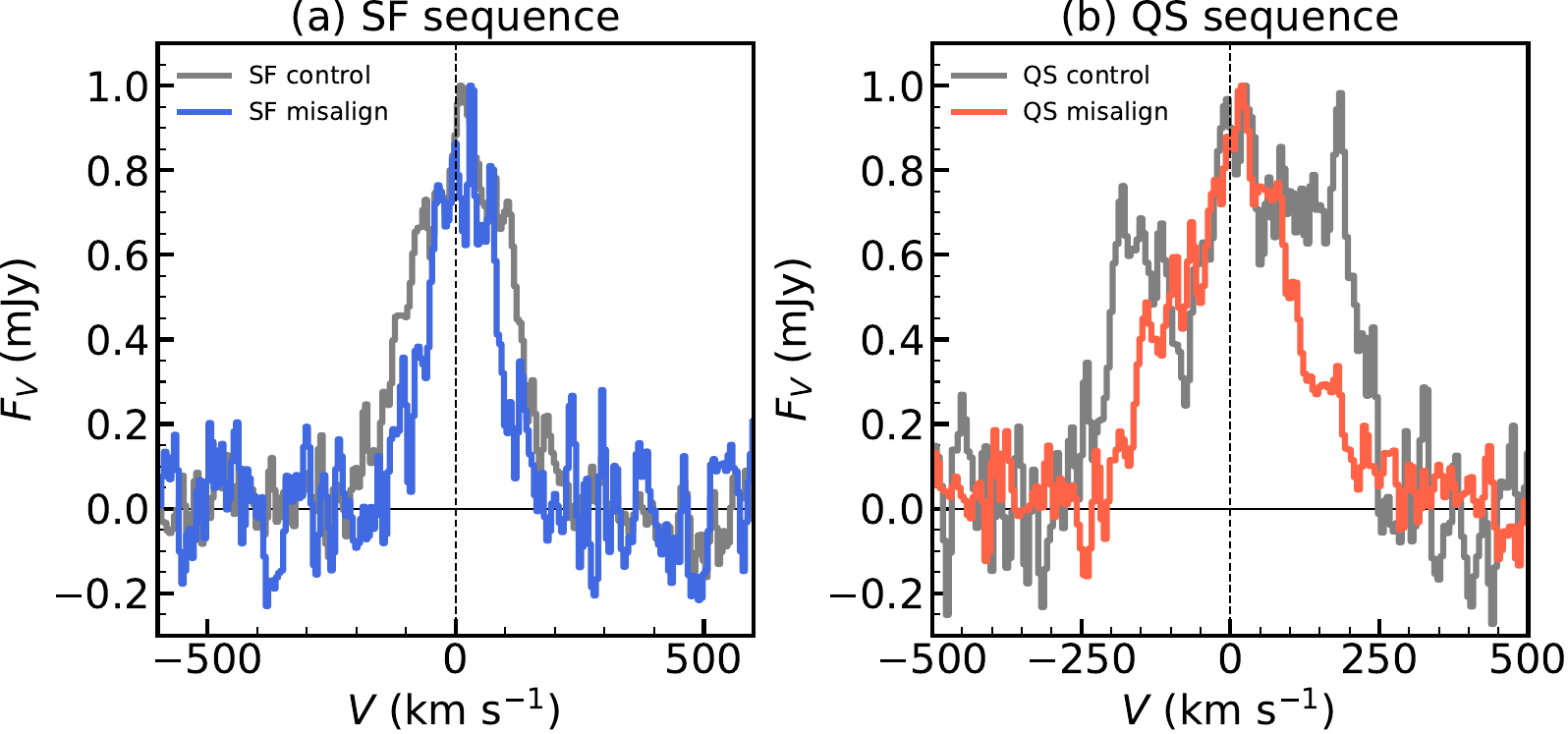}
\caption{Stacked \hi\ spectra in misaligned galaxies and aligned controls. (a) Blue and gray profiles show the stacked \hi\ spectra in SF misaligned galaxies and their aligned controls, respectively. (b) Red  and gray profiles show the stacked \hi\ spectra in QS misaligned galaxies and their aligned controls, respectively. In panels (a) \& (b), the vertical dashed line marks the position of $V = 0~\rm km~s^{-1}$, and the horizontal solid line marks the flux of $F_V = 0~\rm mJy$.}
\label{f:stacking}
\end{figure}

Given the different impact of gas accretion on SF and QS misaligned galaxies, we stack the \hi\ spectra in these sequences separately and make comparisons with corresponding aligned controls. The blue and gray profiles in Figure~\ref{f:stacking}(a) display the stacked \hi\ spectra for SF misaligned galaxies and SF aligned controls. The vertical dashed line marks the position of $V = 0~\rm km~s^{-1}$. The stacked \hi\ spectrum in SF misaligned galaxies ($W_{50} \sim 130.5~\rm km~s^{-1}$) is narrower than that in SF aligned controls ($W_{50} \sim 220.4~\rm km~s^{-1}$).

The red and gray profiles in Figure~\ref{f:stacking}(b) displays the stacked \hi\ spectra for QS misaligned galaxies and QS aligned controls. The line width in QS misaligned galaxies ($W_{50} \sim 231.4~\rm km~s^{-1}$) is obviously narrower than that in QS aligned controls ($W_{50} \sim 406.6~\rm km~s^{-1}$). This tendency is consistent with SF sequence, while the difference is more prominent. As a result of the operation in step (4), all the stacked spectra have positive velocity at the position of peak flux ($V_{\rm peak} > 0~\rm km~s^{-1}$). To quantitatively compare the \hi\ shape and asymmetry between misaligned galaxies and their aligned controls, we measure these parameters for the \hi\ spectrum of each galaxy in the next section.

\section{measurements of \hi\ profile parameters} \label{sec:spectral measurements}

\subsection{CoG method} \label{subsec:method}

\begin{figure}[h]
\centering
\includegraphics[width=0.8\textwidth]{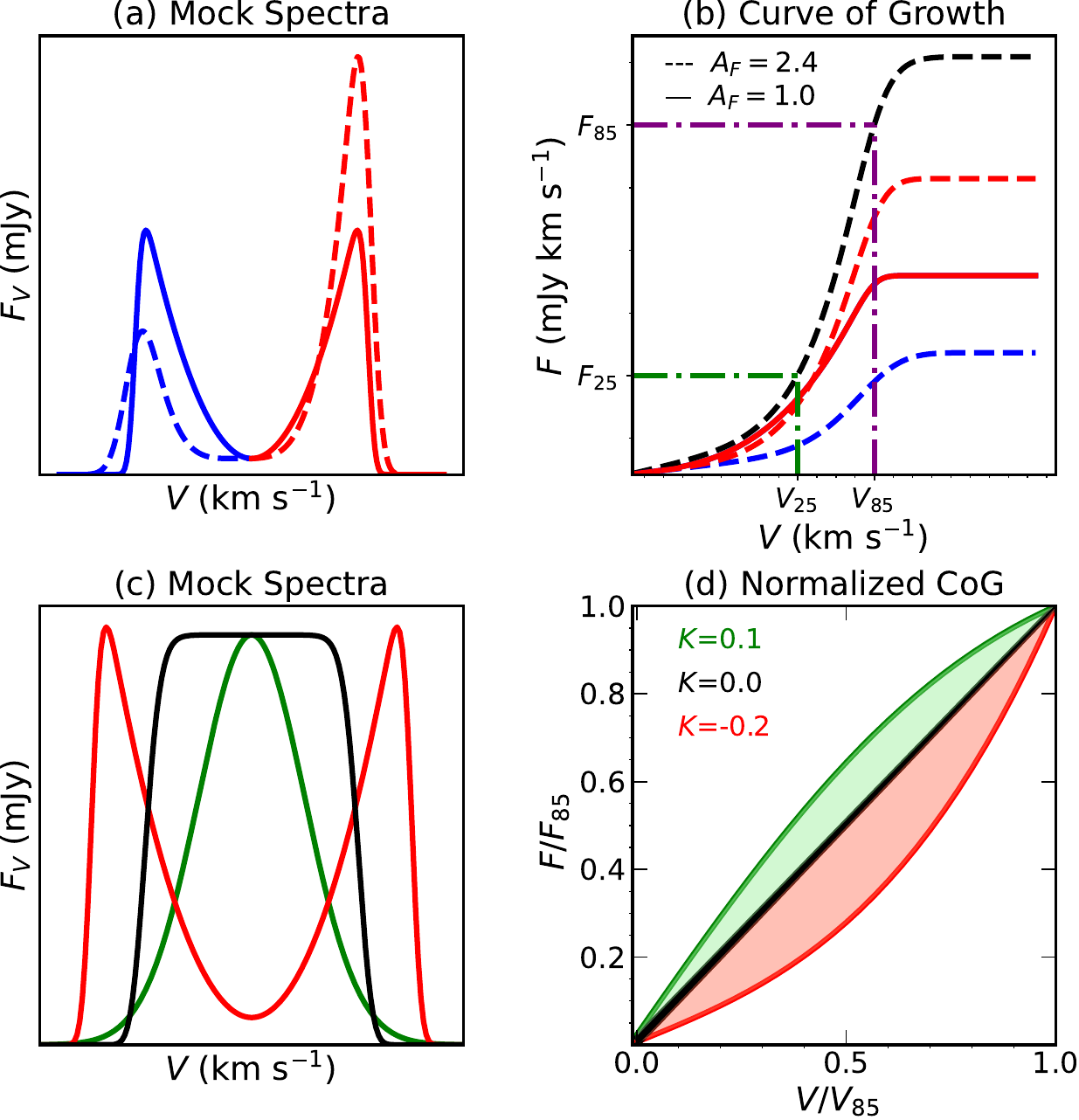}
\caption{Schematic view to illustrate the CoG method. (a) Mock spectra. The solid profile shows a symmetric profile. The dashed profile shows an asymmetric profile. The blue and red colors represent the blueshift and redshift. (b) CoG for mock spectra. The colors and curves are coded the same as panel (a) and the black solid line is the total CoG of solid profile in panel (a). Purple and green lines mark the position of $F_{85}$($V_{85}$) and $F_{25}$($V_{25}$). (c) Mock spectra. The black, green and red profiles display flat-topped, single-peaked and double-horned shapes. (d) Normalized CoG for mock spectra. The colors and curves are coded the same as panel (c). The light-green area represents the $K$ value for the single-peaked profile. The light-red area represents the absolute $K$ value for the double-horned profile.}
\label{f:CoG}
\end{figure}

To measure the shape and asymmetry of each \hi\ spectrum, we apply the curve-of-growth(CoG) method developed by \cite{Yu2020, Yu2022}. For each \hi\ spectrum, we extract channels with flux higher than 0~mJy within the velocity range of $[-500, 500]~\rm km~s^{-1}$. The CoG is built through integrating flux as a function of velocity from line center ($V_{\rm C} = 0~\rm km~s^{-1}$) outward to both blueshifted and redshifted sides. The flux in blueshifted side is integrated within $V\sim[-500, 0]~\rm km~s^{-1}$ ($F_{\rm B}$), while that in redshifted side is integrated within $V\sim[0, 500]~\rm km~s^{-1}$ ($F_{\rm R}$). The $A_{F}$ value indicating \hi\ spectral asymmetry is defined as the larger value of $F_{\rm B}/F_{\rm R}$ and $F_{\rm R}/F_{\rm B}$. Figure~\ref{f:CoG}(a) displays two mock spectra, with blue and red colors representing the blueshifted and redshifted sides, respectively. The solid profile shows a symmetric \hi\ spectrum, while the dashed profile shows an asymmetric \hi\ spectrum. Figure~\ref{f:CoG}(b) displays the CoG on each side of the two spectra, with colors and curves coded in the same way as Figure~\ref{f:CoG}(a). The total CoG for the asymmetric \hi\ spectrum (the dashed spectrum in Figure~\ref{f:CoG}a) is defined as the sum of the integrated flux from the blue-shifted and red-shifted sides, as illustrated by the black dashed line in Figure~\ref{f:CoG}(b). For the symmetric \hi\ spectrum (solid spectrum in Figure~\ref{f:CoG}a), the CoG is exactly the same for both blueshifted and redshifted sides (red solid line in Figure~\ref{f:CoG}b). The $A_{F}$ values for the two \hi\ spectra are labeled in the top-left corner. In general, the higher $A_{F}$ value corresponds to the higher \hi\ asymmetry.

As shown in Figure~\ref{f:CoG}(b), we take 85\% and 25\% of the flux at the flatten part of CoG as $F_{85}$ and $F_{25}$, and define the corresponding velocity widths as $V_{85}$ and $V_{25}$, the concentration of \hi\ profile is characterized as $C_V = V_{85} / V_{25}$. Meanwhile, \cite{Yu2022} introduced a new parameter $K$ to quantify the spectral shape. For each \hi\ spectrum, the CoG is normalized by $V_{85}$ along the velocity axis and by $F_{85}$ along the flux axis. $K$ is then calculated as the area between the normalized CoG and the diagonal line. Figure~\ref{f:CoG}(c) displays three mock spectra, with the black, green and red profiles representing flat-topped, single-peaked and double-horned shapes, respectively. The black profile in Figure~\ref{f:CoG}(d) shows the normalized CoG for flat-topped shape, which exactly follows the diagonal line ($K=0$). The green profile in Figure~\ref{f:CoG}(d) shows the normalized CoG for single-peaked spectral shape, and the light-green area marks the corresponding $K$ value ($K=0.1$). Similarly, the red profile and light-red area show the normalized CoG and absolute $K$ value ($K=-0.2$) for double-horned spectral shape. The $K$ value increasing from negative to positive corresponds for the \hi\ spectral shape changing from double-horned to single-peaked.

\subsection{\hi\ profiles in misaligned galaxies} \label{subsec:analysis}

\begin{figure}[h]
\centering
\includegraphics[width=\textwidth]{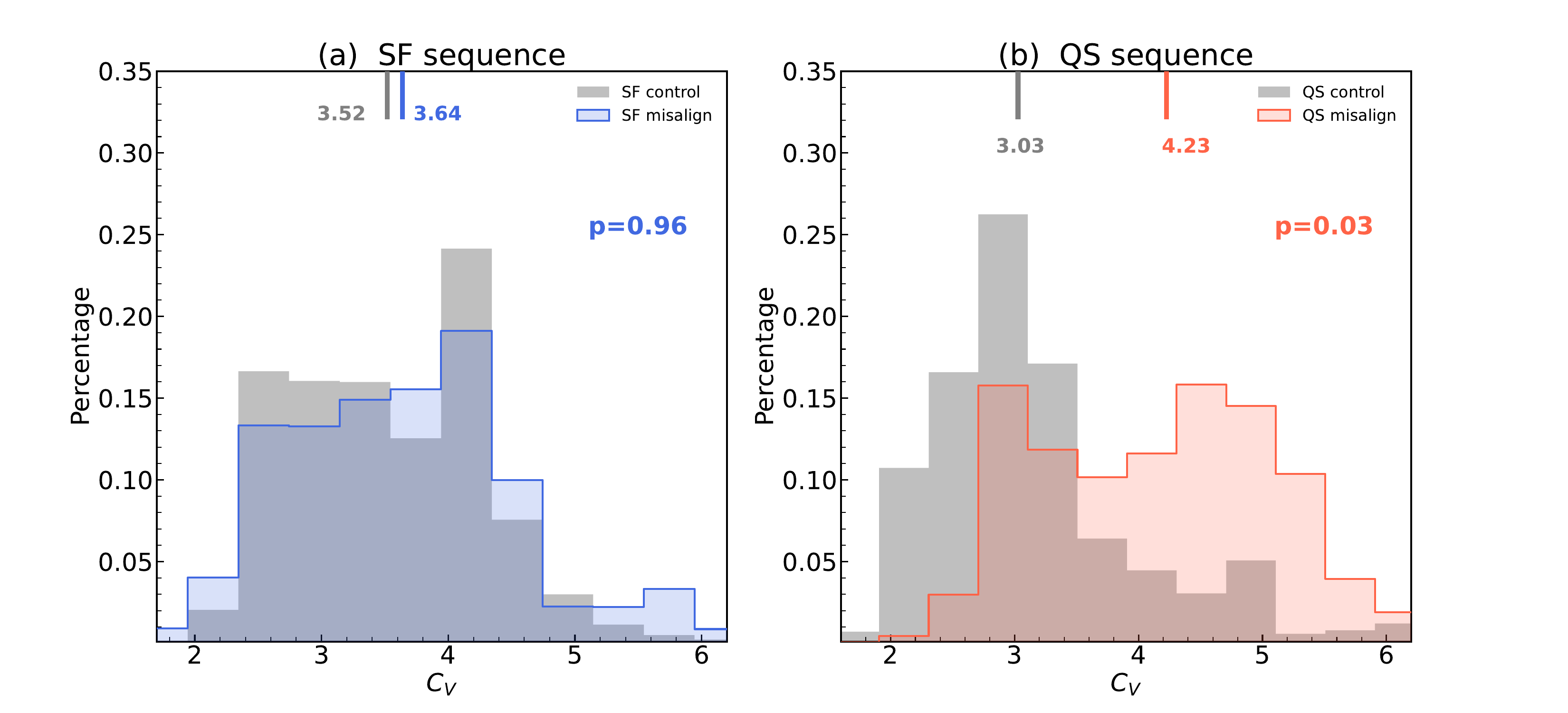}
\caption{The distributions of $C_V$ for misaligned galaxies and aligned controls. (a) The blue and gray histograms show the $C_V$ distribution for SF misaligned galaxies and their aligned controls, respectively. The blue and gray bars on the top mark the corresponding median values. The p-value from KS test of this distribution is marked by blue text in the plot. (b) The red and gray histograms show the $C_V$ distribution for QS misaligned galaxies and their aligned controls, respectively. The red and gray bars on the top mark the corresponding median values. The p-value from KS test of this distribution is marked by red text in the plot.}
\label{f:CVvalue}
\end{figure}

Applying the CoG method to the \hi\ spectra in misaligned galaxies and aligned controls, we can compare the shape of \hi\ profile between them. Figure~\ref{f:CVvalue} displays the distributions of $C_V$. The blue and gray histograms in Figure~\ref{f:CVvalue}(a) show the distributions for misaligned galaxies and aligned controls in SF sequence. The blue and gray bars on the top mark the corresponding median values. The SF misaligned galaxies have similar $C_V$ distribution as their controls with a slightly higher median $C_V$ value. We perform a KS test on these distributions, which is a non-parametric method to quantify the difference between two distributions with p-value $<0.05$ indicating a significant difference existing between two distributions. The KS test gives a p-value of 0.96 for the $C_V$ distributions, which indicates that there is no significant difference in \hi\ profiles between SF misaligned galaxies and aligned controls. The red and gray histograms in Figure~\ref{f:CVvalue}(b) show the distributions of $C_V$ for misaligned galaxies and aligned controls in QS sequence. The bars on the top mark the median values, color-coded in the same way as histograms. The $C_V$ value in QS misaligned galaxies is obviously higher than that in QS aligned controls with a p-value of 0.03. This indicates that the confidence level of difference in $C_V$ distributions between misaligned galaxies and the control sample reaches 97\%, implying more concentrated \hi\ profiles in QS misaligned galaxies than their controls.


\begin{figure}[h]
\centering
\includegraphics[width=\textwidth]{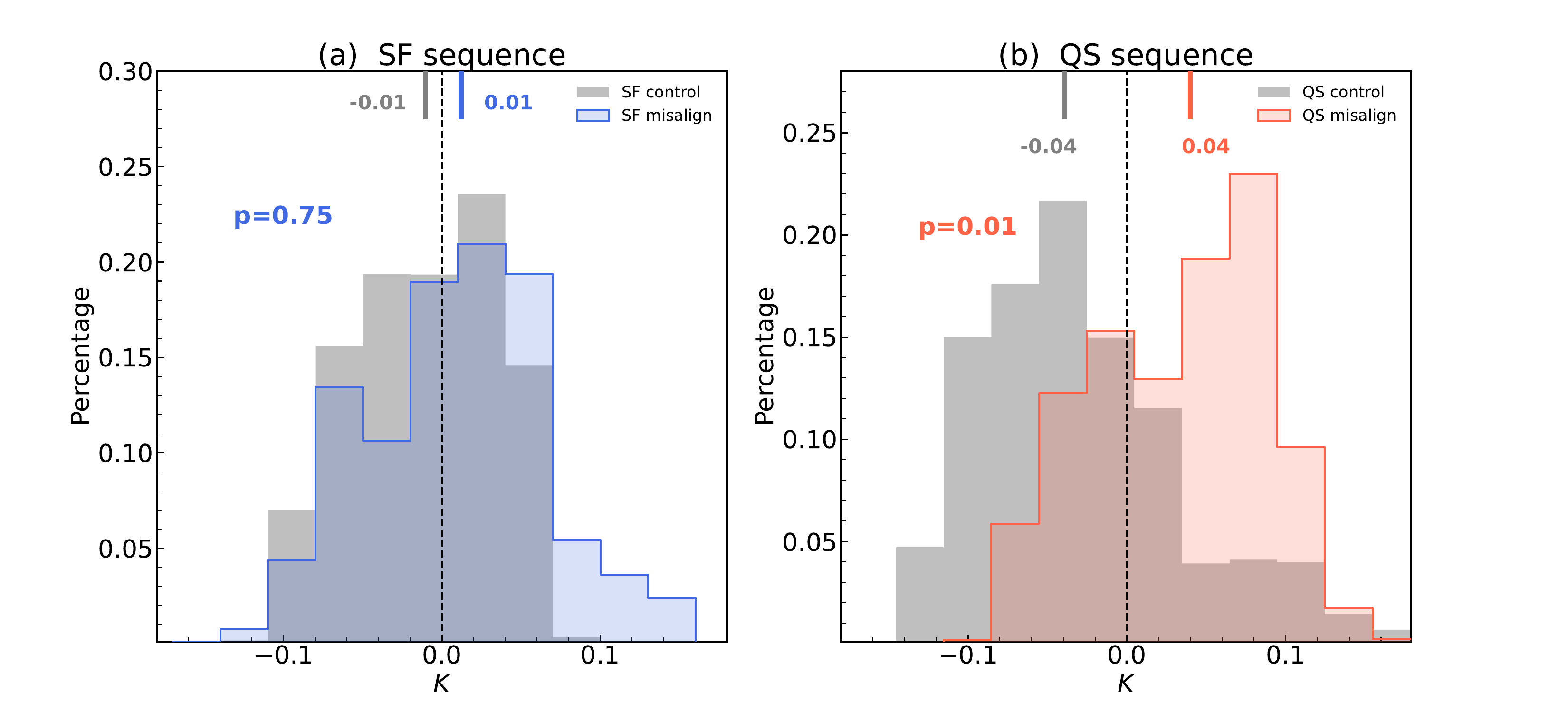}
\caption{The distributions of $K$ for misaligned galaxies and aligned controls. (a) The blue and gray histograms show the $K$ distribution for SF misaligned galaxies and their aligned controls, respectively. The blue and gray bars on the top mark the corresponding median values. The p-value from KS test of this distribution is marked by blue text in the plot. (b) The red and gray histograms show the $K$ distribution for QS misaligned galaxies and their aligned controls, respectively. The red and gray bars on the top mark the corresponding median values. The p-value from KS test of this distribution is marked by red text in the plot. In panels (a) \& (b), the vertical dashed line marks the value of $K = 0$, corresponding to the flat-topped shape.}
\label{f:Kvalue}
\end{figure}

By definition, the two parameters $K$ and $C_V$ are physically related, with higher values of $K$ \& $C_V$ indicating higher concentration of \hi\ profiles.
Since the calculation of parameter $K$ makes use of the full information from the CoG, it can better quantify the \hi\ spectral shape with a positive $K$ value indicating a single-peaked \hi\ profile and a negative $K$ value indicating a double-horned \hi\ profile. On the contrary, $C_V$, relying only on the ratio of two line widths, can only qualitatively measure the concentration of the spectrum. Figure~\ref{f:Kvalue} displays the distributions of $K$ for misaligned galaxies and aligned controls, with colors and symbols coded in the same way as Figure \ref{f:CVvalue}. The vertical dashed line marks the position of $K=0$, corresponding to a flat-topped shape. Figure~\ref{f:Kvalue}(a) displays the $K$ distributions for SF sequence. The SF misaligned galaxies have slightly higher median $K$ value than their controls with a p-value of 0.75. Figure~\ref{f:Kvalue}(b) displays the distributions in QS sequence, where the difference in $K$ between misaligned galaxies and aligned controls is significant. The $K$ value for QS misaligned galaxies tends to be positive, indicating that the \hi\ profiles in these galaxies are dominated by single-peaked profiles. Meanwhile, the $K$ value for QS aligned controls tends to be negative, corresponding to double-horned profiles. KS test gives a p-value of 0.01, indicating that the distributions of parameter $K$ between QS misaligned galaxies and the control sample are different at 99\% confidence level. The \hi\ profiles in QS misaligned galaxies are remarkably more concentrated than that in QS aligned controls.

\begin{figure}[h]
\centering
\includegraphics[width=\textwidth]{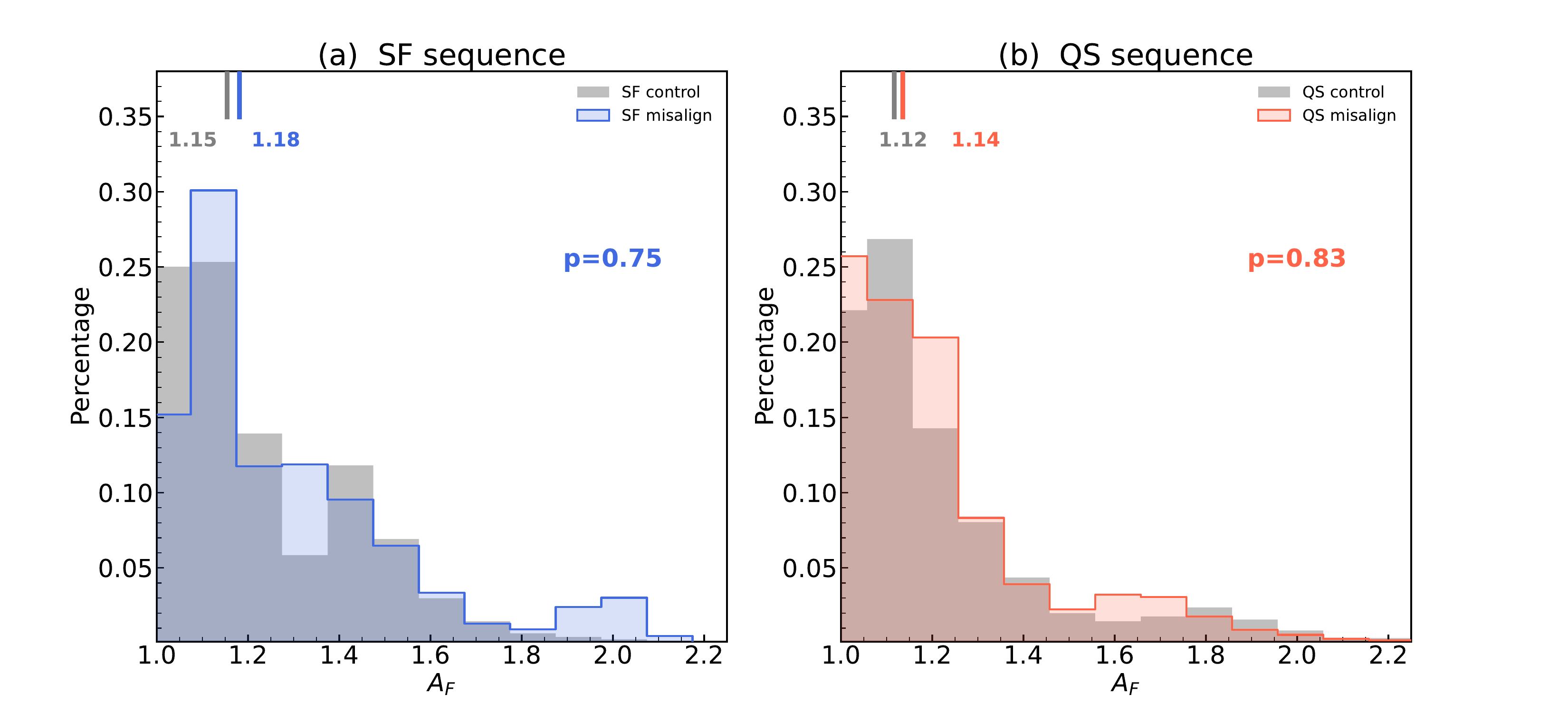}
\caption{The distributions of $A_F$ for misaligned galaxies and aligned controls. (a) The blue and gray histograms show the $A_F$ distribution for SF misaligned galaxies and their aligned controls, respectively. The blue and gray bars on the top mark the corresponding median values. The p-value from KS test of this distribution is marked by blue text in the plot. (b) The red and gray histograms show the $A_F$ distribution for QS misaligned galaxies and their aligned controls, respectively. The red and gray bars on the top mark the corresponding median values. The p-value from KS test of this distribution is marked by red text in the plot.}
\label{f:Asymmetry}
\end{figure}

Figure \ref{f:Asymmetry} displays the distributions of the asymmetry parameter $A_F$, with colors and symbols coded in the same way as Figure~\ref{f:CVvalue}. It is obvious that the difference in $A_F$ distributions between misaligned galaxies and aligned controls can be neglected in both SF sequence (Figure~\ref{f:Asymmetry}a) and QS sequence (Figure~\ref{f:Asymmetry}b) with a p-value of 0.75 (0.83) for SF (QS) sequence. Although the \hi\ spectra we analyzed are single-dish without any spatial resolution, the similar distributions of the parameter $A_F$ in misaligned galaxies and their aligned controls imply the consistent spectral asymmetry of \hi\ spectra in these galaxies and that external gas accretion does not lead to significant asymmetry in gas kinematics. \cite{Zhou2022} compared the asymmetry of the ionized gaseous velocity fields between misaligned galaxies and their aligned controls selected from the MaNGA survey, finding obviously higher asymmetry for misaligned galaxies in both SF sequence and QS sequence. The inconsistency between the results from the current work and that from \cite{Zhou2022} might be due to that we study different gas-phases, \hi\ traced cold gas versus optical emission lines traced ionized gas in \cite{Zhou2022}. Without further spatially resolved observations in \hi\ gas, we will not over-interpret this difference in asymmetry between different gas phases here.

\section{Impact of external gas accretion on \hi\ profile} \label{sec:impact}
The distributions of \hi\ line widths in misaligned galaxies and their aligned controls (shown in Figure~\ref{f:linewidth distribution}) indicate that misaligned galaxies exhibit statistically narrower line widths compared with their aligned controls. Additionally, the $C_V$ in Figure~\ref{f:CVvalue}, a parameter traditionally used to indicate the spectral concentration, shows a consistent trend with the results for the superior parameter $K$ in Figure~\ref{f:Kvalue}. These findings, based on individual \hi\ line measurements, are consistent with the comparison of stacked spectra in Figure~\ref{f:stacking}.
On the one hand, for both SF sequence and QS sequence, the misaligned galaxies show higher $K$ (or $C_V$) values, representing more concentrated \hi\ profiles. On the other hand, the difference in $K$ (or $C_V$) values between misaligned galaxies and aligned controls is more prominent in QS sequence compared to the SF sequence, indicating more obvious influence of \hi\ gas accretion on QS misaligned galaxies than SF ones.

Previous works have done a lot of studies based on the optical spectra and suggested that the impact of external gas accretion on SF and QS misaligned galaxies is different. For SF misaligned galaxies, both \cite{Chen2016} and \cite{Jin2016} found lower D$_n4000$ (younger stellar population) and higher sSFR in the central regions than their outskirts. \cite{Xu2022} found positive D$_n4000$ gradients in SF misaligned galaxies while negative gradients in aligned controls. These results can be explained by a picture that the progenitors of SF misaligned galaxies accrete misaligned gas from gas-rich dwarfs or cosmic web. The interaction between accreted gas and pre-existing gas leads to the re-distribution of angular momentum, causing large amount of gas inflow and triggering the central star formation. For QS galaxies, the lack of difference in stellar population gradients between misaligned galaxies and aligned controls is due to that the progenitors of QS misaligned galaxies are gas-poor, the angular momentum loss mechanism described above does not exist in the QS misaligned galaxies \citep{Zhou2022}.

\begin{figure}[h]
\centering
\includegraphics[width=\textwidth]{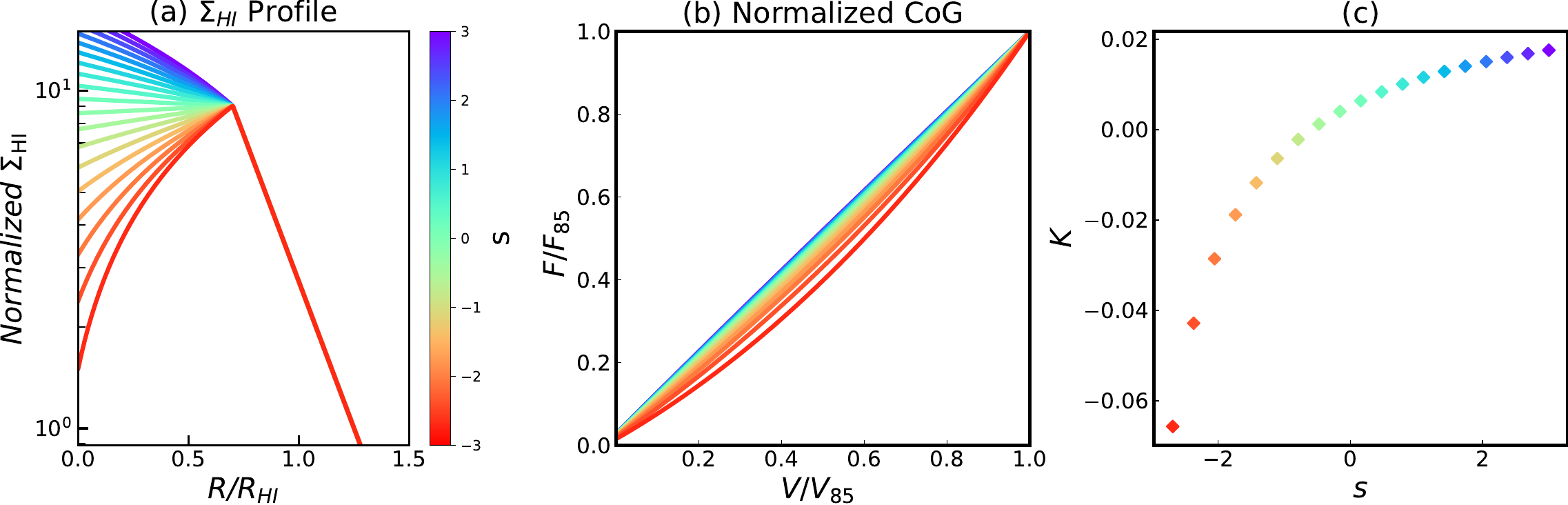}
\caption{The variations of the $K$ values with different \hi\ surface density profiles. (a) \hi\ surface density profiles, with the colorbar showing the $s$ values. From bottom to top, the \hi\ surface density at the center of the galaxy goes from deficient to enriched responding from the red to the purple solid lines. (b) The normalized CoG obtained from the spectra corresponding to each \hi\ surface density distribution in panel (a) with all other parameters fixed, and the area enclosed by the CoG and the diagonal line is the corresponding $K$ value with the color-coded in the same way as panel (a). (c) Relationship between \hi\ surface density profiles and $K$ values. $s$ represents the slope of \hi\ surface density in galactic center with a larger $s$ indicating a higher central \hi\ surface density. The color-code is in the same way as panel (a).}
\label{f:sigma}
\end{figure}

Based on this picture, we would expect that the difference in \hi\ spectral shape should be more apparent in SF misaligned galaxies and their aligned controls. However, our observational results clearly show significant \hi\ spectral shape differences in QS misaligned galaxies and their aligned controls, the difference is not obvious in SF misaligned galaxies and their aligned controls. To understand this result, we simulate variations of the $K$ values measured by the CoG method for galaxies with different \hi\ surface density distributions. Figure~\ref{f:sigma}(a) displays the mimic \hi\ surface density ($\Sigma_{\rm HI}$) profiles, which can be described as segmented functions. For $r<0.7R_{\rm HI}$, $\Sigma_{\rm HI} \propto -s(R/R_{\rm HI})+b$, with $R_{\rm HI}$ representing the radius of a galaxy where averaged $\Sigma_{\rm HI}\sim1~M_\odot/\rm pc^2$. The $s$ values ranging from negative (central \hi-deficient) to positive (central \hi-enriched) corresponds to the profiles from red to purple, as shown in the colorbar of Figure~\ref{f:sigma} (a). At $r>0.7R_{\rm HI}$, $\Sigma_{\rm HI}$ is described as an exponential decline function $\Sigma_{\rm HI} \propto e^{-R/R_{\rm HI}}$ \citep{Wang2016}. Combining the $\Sigma_{\rm HI}$ profiles with a uniform rotation curve, we can have the \hi\ emission profile models. In Figure~\ref{f:sigma}(b) we show the normalized CoG measured from \hi\ emission profile models and the derived $K$ in Figure~\ref{f:sigma}(c). Different colors in Figure~\ref{f:sigma}(b) and (c) are corresponds to different $\Sigma_{\rm HI}$ profiles shown in Figure~\ref{f:sigma}(a). As shown in Figure~\ref{f:sigma}(c), the dependence of $K$ on $s$ can be described by a double power law. For $s < 0$, $K$ increases with $s$ very quickly, namely, as a galaxy transits from central \hi-deficient to central \hi-enriched (from red curve to cyan curve in Figure~\ref{f:sigma}a), its spectrum changes from double-horned to single-peaked shape, accompanied by a significant increase in $K$ value from negative to positive. However, at $s > 0$, the dependence weakens, $K$ increases slowly with $s$, namely, once the galaxy is already central \hi-enriched (from cyan curve to purple curve in Figure~\ref{f:sigma}a), further increases in central \hi\ density will not have strong impact on measured $K$ values. \cite{Wang2016} compared the radial gradients of $\Sigma_{\rm HI}$ in different types of galaxies, and found that early-type galaxies have lower $\Sigma_{\rm HI}$ in the inner region, whereas spiral galaxies show an exponential decline of $\Sigma_{\rm HI}$ radial profile. \cite{Bigiel2008} identified a universal saturation threshold ($\Sigma_{\rm HI}\sim9~M_\odot/\rm pc^2$) for atomic hydrogen (\hi) surface density. They suggest that \hi\ gas is effectively converted into $\rm H_2$ at $\Sigma_{\rm HI}>9~M_\odot/\rm pc^2$, providing fuel for the formation of new stars. This saturation is observed in galactic centers, outskirts, and dwarf galaxies.
As a consequence, the difference in $K$ and $C_V$ between misaligned galaxies and their aligned controls can be explained in the following ways:
\begin{itemize}
\item[1.]The progenitors of SF misaligned galaxies have central enriched \hi\ ($s>0$), which leads to single-peaked \hi\ emission profiles. The interaction between accreted \hi\ and pre-existing \hi\ re-distributes the AM of \hi\ gas, inducing gas inflow (as suggested by \citealt{Chen2016}). However, the lack of obvious difference in $K$ and $C_V$ values between misaligned galaxies and their aligned controls is due to the insensitivity of $K$ and $C_V$ on $\Sigma_{\rm HI}$ gradient at $s>0$ (as simulated in Figure~\ref{f:sigma}). Another possible scenario is that the inflowing \hi\ gas in SF misaligned galaxies is converted into molecular gas and trigger star formation, preventing significant increase of \hi\ gas surface density in their central regions.
\item[2.]The progenitors of QS misaligned galaxies are dominated by early-type galaxies, they are initially central \hi-deficient ($s<0$) with double-horned \hi\ profiles. On the one hand, the interaction between the accreted \hi\ and the pre-existing \hi\ leads to the re-distribution of gas AM and induces gas inflow, replenishing central \hi\ reservoirs. The transition from central \hi\ deficiency to sufficiency---a `from-scratch' replenishment process---causes a significant increase in $K$ or $C_V$ values for misaligned galaxies. On the other hand, although these QS misaligned galaxies have undergone cold gas inflow processes, the \hi\ gas surface densities in their central regions remain below the critical threshold ($\sim 9~M_\odot/\rm pc^2$) required for \hi-$\rm H_2$ transition. This is consistent with the results from optical observations in \cite{Zhou2022}, where they did not find any new formed stars in the QS misaligned galaxies.
\end{itemize}


\section{Conclusion} \label{sec:conclusion}

In this study, we build a sample of 37 gas-star misaligned galaxies with robust \hi\ observations from the final data release of the MaNGA. To explore the impact of external gas accretion on the \hi\ distribution, we construct a control sample of aligned galaxies with similar stellar mass, D$_n$4000 and inclination angle as well as robust \hi\ measurement. We also classify the misaligned galaxies and aligned controls into star-forming (SF) sequence and quiescent (QS) sequence to investigate the impact separately. The main observational results are as follows:

\begin{itemize}

    \item[1.] The stacked \hi\ spectra in misaligned galaxies are narrower in line width than their aligned controls in SF sequence and QS sequence. The difference in line width is more obvious in the QS sequence.

    \item[2.] The median $C_V$ and $K$ values in SF misaligned galaxies are slightly higher than SF aligned controls, while the distributions of them are comparable. The $C_V$ and $K$ distributions in QS misaligned galaxies are prominently higher than QS align controls, with \hi\ spectra are dominated by single-peaked shapes in the misaligned galaxies and double-horned shapes in the aligned controls.

    \item[3.] The difference in \hi\ spectral asymmetry between misaligned galaxies and aligned controls can be neglected in both SF sequence and QS sequence. Without further spatially resolved observations in \hi, we will not over-interpret this difference in this work.


\end{itemize}

The difference in spectral shape of misaligned galaxies and their aligned controls suggests that the interaction between the external accreted gas and the pre-existing gas leads to the re-distribution of gas angular momentum and induces gas inflow. The progenitors of SF misaligned galaxies are central \hi-enriched. In this case, the measured $K$ value is not sensitive to the increase of central \hi\ surface density. On the other hand, as the \hi\ surface density increases to $\sim 9~M_\odot/\rm pc^2$, the \hi-$\rm H_2$ transition happens, preventing significant increase of \hi\ gas surface density in their central regions. The progenitors of QS misaligned galaxies are central \hi-deficient with double-horned \hi\ profiles. The transition from central \hi-deficient to central \hi-enriched through external gas accretion results in a significant increase in $K$ or $C_V$ values for misaligned galaxies. Although these QS misaligned galaxies have undergone cold gas inflow processes, the central \hi\ gas surface densities fail to reach the critical threshold for \hi-$\rm H_2$ transition ($\sim 9~M_\odot/\rm pc^2$). This finding agrees with the results from optical observations in \cite{Zhou2022}, which found no evidence of newly formed stars in QS misaligned galaxies.

\textbf{Acknowledgements:} MB acknowledges support by the National Natural Science Foundation of China, NSFC Grant no. 12303009. YMC acknowledges support by the National Natural Science Foundation of China, NSFC Grant no. 12333002 and the China Manned Space Project, no. CMS-CSST-2025-A08. LCH was supported by the National Science Foundation of China, NSFC Grant no. 12233001 and the National Key R\&D Program of China (2022YFF0503401). NKY acknowledges support by the projects funded by China Postdoctoral Science Foundation no. 2022M723175 and GZB20230766.

Funding for the Sloan Digital Sky Survey IV has been provided by the Alfred P. Sloan Foundation, the U.S. Department of Energy Office of Science, and the Participating Institutions. SDSS- IV acknowledges support and resources from the Center for High-Performance Computing at the University of Utah. The SDSS web site is www.sdss.org. SDSS-IV is managed by the Astrophysical Research Consortium for the Participating Institutions of the SDSS Collaboration including the Brazilian Participation Group, the Carnegie Institution for Science, Carnegie Mellon University, the Chilean Participation Group, the French Participation Group, Harvard-Smithsonian Center for Astrophysics, Instituto de Astrof\'{i}sica de Canarias, The Johns Hopkins University, Kavli Institute for the Physics and Mathematics of the Universe (IPMU) / University of Tokyo, Lawrence Berkeley National Laboratory, Leibniz Institut  f\"{u}r Astrophysik Potsdam (AIP), Max-Planck-Institut  f\"{u}r   Astronomie  (MPIA  Heidelberg), Max-Planck-Institut   f\"{u}r   Astrophysik  (MPA   Garching), Max-Planck-Institut f\"{u}r Extraterrestrische Physik (MPE), National Astronomical Observatory of China, New Mexico State University, New York University, University of Notre Dame, Observat\'{o}rio Nacional / MCTI, The Ohio State University, Pennsylvania State University, Shanghai Astronomical Observatory, United Kingdom Participation Group, Universidad Nacional  Aut\'{o}noma de M\'{e}xico,  University of Arizona, University of Colorado  Boulder, University of Oxford, University of Portsmouth, University of Utah, University of Virginia, University  of Washington,  University of  Wisconsin, Vanderbilt University, and Yale University.

\bibliography{sample631}{}
\bibliographystyle{aasjournal}

\end{document}